# Retention Horizons of Volatile Stores in iOS and Android Extractions

Daniel Kwaku Ntiamoah Addai
SectraTEK Lab, Prudential Associates
daniel@prudentialassociates.com
*Corresponding author*

Sanders Larweh Tetteh
Booking.com
sanderstetteh@gmail.com

Selorm Adjabeng
Cyberteq Falcon Ltd
sirlormselorm@gmail.com

Kingsley Albert Egyir
Northeastern University
egyir.k@northeastern.edu

**Abstract.** Whether an extraction can answer a question about a given day depends on how far back each store reaches, yet public statements rest on single test devices. We measured the retention horizon, the time from a store's oldest retained record to acquisition, for eight iOS and seven Android stores on 48 full file system extractions and seven bug reports, keeping devices whose clock agreed with the acquisition time (26 Apple devices, 13 Android phones). Horizons ranged from minutes to years and followed the mechanism that bounds each store. The unified log Persist stream reached a median of 2.5 days; interactionC reached 166 to 170 days on 18 of 25 devices. Android stores stayed within their source-code limits (usage statistics 10.2 days, DropBox 3.1, notifications 1.5); bug-report logs reached 3.1 hours at most. The oldest record misled three times: one knowledgeC stream reached back two years, the others a median of 29 days; interactionC read without its write-ahead log showed older records on 18 of 25 devices; and four of seven bug reports began with stray lines up to 16.6 days old above a body of 3.1 hours or less. A horizon belongs to a store, a release and a device.

## 1 Introduction

Many records on a phone are kept only for a while. Log streams are trimmed when they reach a size, usage databases drop rows older than a set age, and diagnostic folders keep a fixed number of files. For an examiner the consequence is simple: a question about a day can be answered only if that day still lies inside the window of the stores that would hold the answer. We call that window the *retention horizon* of a store: the time between its oldest retained record and the acquisition.

**A motivating example.** Studies of insider theft of intellectual property found, in one SEI report, that many insiders who stole intellectual property did so within 30 days of their termination [15], so an examiner may be asked what a phone did in that month. On the 19 iOS 26 devices in our sample whose clock we could check, the Persist stream of the unified log reached back 30 days on none (0 of 19; Wilson 95% interval 0 to 17%), the Special stream on 7, and the CoreDuet people-interaction store (interactionC) on all 19 (Section 5.6).

**The problem.** Public statements about these windows are approximate and come from one test device or from no stated measurement: unified logs last hours to about ten days and a sysdiagnose holds little from previous days [4], unified-log data are flushed within a few days [11], knowledgeC holds about four weeks [10], and knowledgeC and Biome data expire after 28 to 30 days [13] (Section 2). For Android, the platform source states limits for several stores [5]–[9], but whether phones in use reach those limits, and how much horizons vary between devices on the same release, has not been measured as far as we found. We ask four research questions.

- **RQ1 (Horizons).** What is the distribution of retention horizons of the main volatile stores on iOS and Android extractions and in Android bug reports?
- **RQ2 (Variation).** How much do horizons vary between devices on the same OS release?
- **RQ3 (Coverage).** Which stores cover the 30 days before acquisition?
- **RQ4 (Measurement).** Does a store's oldest record measure the window an examiner can use?

**Contributions.** This report makes four contributions.

- A pre-stated measurement protocol for retention horizons: a fixed store list, one dating rule per store, an acquisition time, and a clock rule that admits a device only when its newest record agrees with the acquisition time (Section 3).
- Horizon distributions for 15 stores on 26 Apple devices and 13 Android phones and for two log sections of seven bug reports, with medians, order-statistic intervals and cells of at least five devices (Sections 5.2 to 5.4).
- A comparison of measured Android horizons with the limits in the platform source, and of iOS horizons with published statements (Sections 5.3 and 6.1).
- Evidence that the oldest record can mislead, through one long-lived knowledgeC stream, through records that the write-ahead log has already deleted, and through stray lines above the body of a bug-report log (Sections 5.4, 5.7 and 5.8).

> **Scope.** All results describe the Lab's research corpus: 48 full file system extractions and seven bug reports readable when the study ran, of which 39 devices and seven reports passed the clock rule. Most cells are iOS 26 or Android 16, and Samsung supplies 11 of the 13 Android phones; cells of fewer than five devices are not reported. A horizon measures what a device still held at acquisition, which depends on the store's limit, on how much was written and on how long the device had been in use. Section 7 states the limits.

## 2 Background and Related Work

### 2.1 Volatile stores and how they expire

Table 2 lists the stores measured, where each is kept, the field that dates it, and the limit stated in public sources. On iOS, the unified log keeps compressed `tracev3` files in several streams under `/private/var/db/diagnostics`. Each file begins with a header chunk that holds the wall-clock time at which the file was started, in seconds since 1970 UTC [14]. On macOS, the Persist folder is managed to a total size of about 520 to 530 MB, while Special entries are removed by type and content rather than by size [16]; we found no equivalent statement for iOS. The CoreDuet databases `knowledgeC.db` and `interactionC.db` hold device and app events and interactions with people. Biome keeps event streams in segment files; we read the numeric name of a segment as its creation time in microseconds since 2001, a reading we checked against file times (Section 3.5). Powerlog archives are daily compressed copies of the power log database named by calendar date.

On Android, the usage statistics service writes one file per day, week, month and year, named by the start time of the period. The notification history, DropBox, network statistics and process statistics services each keep files named by a time. A bug report carries the system and event log buffers as text sections whose lines have a month, a day and a time but no year.

### 2.2 Practitioner statements on retention

Afonin [4] examined the logs of one iOS test device. He reports that most unified-log entries expire within minutes or hours, that a log collection covered up to the last ten days, that a sysdiagnose contained little from previous days, and that the view that logs are kept for 30 days is largely incorrect; shutdown and Wi-Fi logs went back to the device's activation. A 2020 practitioner post on AirDrop records in the unified log notes that the data are flushed quickly and may be available for only a few days [11]. For knowledgeC, an often-cited 2018 description says that its object table holds usage entries for approximately four weeks [10], and a 2024 vendor post lists knowledgeC and Biome data among sources that expire after 28 to 30 days, without stating a method or sample [13]. For macOS, Oakley describes the size target of the Persist folder and the separate purging of Special entries [16]. On Android, the intrusion log introduced with Android 16 is described as kept for 12 months in the user's account; the description concerns Pixel devices [12], and no device in our sample carried it.

### 2.3 Platform source code

The Android Open Source Project states limits for several stores in code (Table 2). The usage statistics database keeps at most 100 daily, 50 weekly, 12 monthly and 10 yearly files and prunes daily files older than 10 days, weekly older than four weeks, monthly older than six months and yearly older than two years [9]. Notification history is pruned after one day [7]. DropBox entries have a default maximum age of three days, a default of 1,000 files and a size quota, each of which a global setting can change [5]. Network statistics files cover 15 days and are deleted after 90 days for device and per-app data [6]. Process statistics keep at most eight committed state files [8]. These are limits; a phone reaches them only if it has run long enough and no vendor setting overrides them.

### 2.4 Earlier reports of this series

Three earlier reports touch retention for one store each. LAB-TR-0003 described the App Ops history databases of Android 16 and the time windows they store [1]. LAB-TR-0004 showed that the clock behind a date must be checked before the date is used, and supplies the clock discipline of Section 3.3 [2]. LAB-TR-0009 found that the Screen Time usage tables were mostly empty and that counts changed when the write-ahead log was not applied [3]; we therefore did not include Screen Time here, and we read every database twice, with and without its log.

**Table 1.** Research questions, answers and evidence. Days are counted back from acquisition; all figures describe clock-checked devices of this sample.

| RQ | Answer | Evidence |
|---|---|---|
| RQ1 | From hours (bug-report logs, at most 3.1 h) through days (unified log Persist, median 2.5 d; Android daily usage statistics, median 9.8 d) to months (interactionC, median 168 d; Android network statistics, median 94 d). | Tables 4, 5; Fig. 2 |
| RQ2 | Wide for stores bounded by size or file count (Persist on iOS 26: 0.39 to 21.2 d; process statistics on Android 16: 1.7 to 46.2 d); narrow for stores bounded by age (interactionC on iOS 26: quartiles 167.0 to 168.1 d). | Table 6 |
| RQ3 | iOS: interactionC (25 of 25) and Biome App.InFocus (20 of 22); the Special log stream on 11 of 25, Persist on 1 of 25. Android: network statistics (12 of 13) and coarse usage aggregates; no daily usage, DropBox or notification store; no bug-report log. | Fig. 3 |
| RQ4 | Not always. One knowledgeC stream sets the oldest record on 22 of 26 devices; without it the median horizon falls from 157 to 29 d. interactionC without its write-ahead log shows an older oldest record on 18 of 25 devices. In 4 of 7 bug reports the oldest system-log lines are stray lines up to 16.6 d older than a body of at most 3.1 h. | Sections 5.4, 5.7, 5.8 |

**Position of this work.** The statements above give single values or ranges from one device, or source-code limits that say what a store may hold, not what devices in use held. We found no measurement across many devices of how far back these stores reach, no distribution within one release, and no treatment of the clock behind the oldest record. The stores, their paths and the Android limits are known from the cited sources; the horizon distributions, the within-release variation, the comparison of measured and stated values, and the three ways in which the oldest record misled are new here, as far as we found.

# 3 Method

The store list, strata, thresholds and statistics were written into a dated note before any horizon was computed; the analyses marked *post hoc* in Section 5 were added after the first results.

## 3.1 Definitions

For a store $S$ on a device, let $T$ be the acquisition time, $o$ the time of the oldest retained record and $n$ the time of the newest. The *retention horizon* is $T - o$ and the *newest lag* is $T - n$, both in days. A record is a database row, a log line, or a file whose name or header carries a time. For record-level stores we also report the horizon of the first percentile of record times, which is robust to a few stray old records. For the log sections of a bug report the oldest line is not used, because a few stray old lines can sit far above the rest of the section (Section 5.4). The section is split at every gap of ten minutes or more between consecutive line times. The lines before the last such gap are *head strays* if they number at most 0.5% of the lines of the section (and at least one line); the *body* then starts at the first line after that gap, and the horizon of the log is $T$ minus the time of that line. If no gap qualifies, the body starts at the oldest line. The two constants were chosen before the body horizons were computed and are tested in Section 5.4; the age of the oldest line is reported alongside.

### 3.2 Dating each store

Every store is dated by one field (Table 2). For unified-log streams we read the first 48 bytes of each `tracev3` file and took the wall-clock start time at offset 32 of the header chunk [14]; a file's first record lies at or shortly after its start, so the oldest file start can overstate the horizon only by the delay before that first write. For the CoreDuet databases we copied the database and its `-wal` file to a private scratch area and read the minimum and maximum start dates with SQL, once with the log applied and once from the main file alone. Biome segments, powerlog archives and the Android stores were dated from their file names. Two names carry device local time with no zone (process statistics and Samsung battery history), and powerlog archive names carry a calendar day; their horizons carry an uncertainty of up to 14 h and one day respectively, which does not change a horizon measured in days. In bug reports we read the `SYSTEM LOG` and `EVENT LOG` sections, took the year from the report's header time, and assigned the previous year to any line that would otherwise fall more than one day after the header. Horizons of these sections are counted from the start of the body (Section 3.1).

### 3.3 Acquisition time and the clock rule

The acquisition time was the end time in the extraction record kept with the archive, converted with its UTC offset; where no such record existed it was the modification time of the archive file on the workstation, a weaker reference because copying can change it. A device's horizons were used only if its clock passed the rule in Listing 1. The rule compares the newest record of the measured stores, written by the device clock, with the acquisition time, written by the workstation clock. A device in use at acquisition writes records until the copy starts, so a large gap indicates a wrong device clock, a wrong acquisition time, or a device that had stopped writing; the rule cannot tell these apart and excludes all three. Records dated before 2007 or after the acquisition were counted and excluded from the oldest record. For a bug report both times come from one device clock, so a constant clock offset cancels; we counted backward steps of more than one hour in line order as evidence of a clock change within the window.

**Table 2.** Stores measured, how each is dated, and the retention stated in public sources. Paths are shortened; "name" means the time is read from the file name.

| Store | Where kept | Dating field and clock | Stated retention |
|---|---|---|---|
| ***iOS full file system*** | | | |
| Unified log: Persist, Special, Signpost, HighVolume | db/diagnostics/<stream>/*.tracev3 | File header start time, UTC [14] | Hours to about 10 days [4]; a few days [11] |
| knowledgeC | CoreDuet/Knowledge/knowledgeC.db | ZOBJECT.ZSTARTDATE, seconds since 2001 | About 4 weeks [10]; 28 to 30 days [13] |
| interactionC | CoreDuet/People/interactionC.db | ZINTERACTIONS.ZSTARTDATE | None found |
| Biome App.InFocus | Biome/streams/restricted/App.InFocus/local/ | Segment name, microseconds since 2001 | 28 to 30 days (Biome data) [13] |
| Powerlog archives | BatteryLife/Archives/powerlog_<date>_* | Calendar date in name (day, ±1 d) | None found |
| ***Android full file system (user 0)*** | | | |
| Usage statistics | usagestats/{daily, weekly, monthly, yearly}/ | Name, epoch ms | Source: daily 10 d, weekly 4 wk, monthly 6 mo, yearly 2 y [9] |
| Notification history | notification_history/history/ | Name, epoch ms | Source: 1 d [7] |
| DropBox entries | system/dropbox/<tag>@<ms>.* | Name, epoch ms | Source default: 3 d, 1,000 files, a size quota [5] |
| Network statistics | netstats/{xt, uid, uid_tag}.<start>-<end> | Name (start), epoch ms | Source: files of 15 d, deleted after 90 d (device and per-app) [6] |
| Process statistics | procstats/state-v2-<date>.bin | Name, device local time (±14 h) | Source: 8 files [8] |
| Battery history (Samsung) | data/log/batterystats/ | Name, device local time (±14 h) | None found |
| ***Android bug report*** | | | |
| System log, event log | SYSTEM LOG, EVENT LOG sections | Line time, counted from the start of the body of the section (Section 3.1), against the report's header time; one device clock | None found |

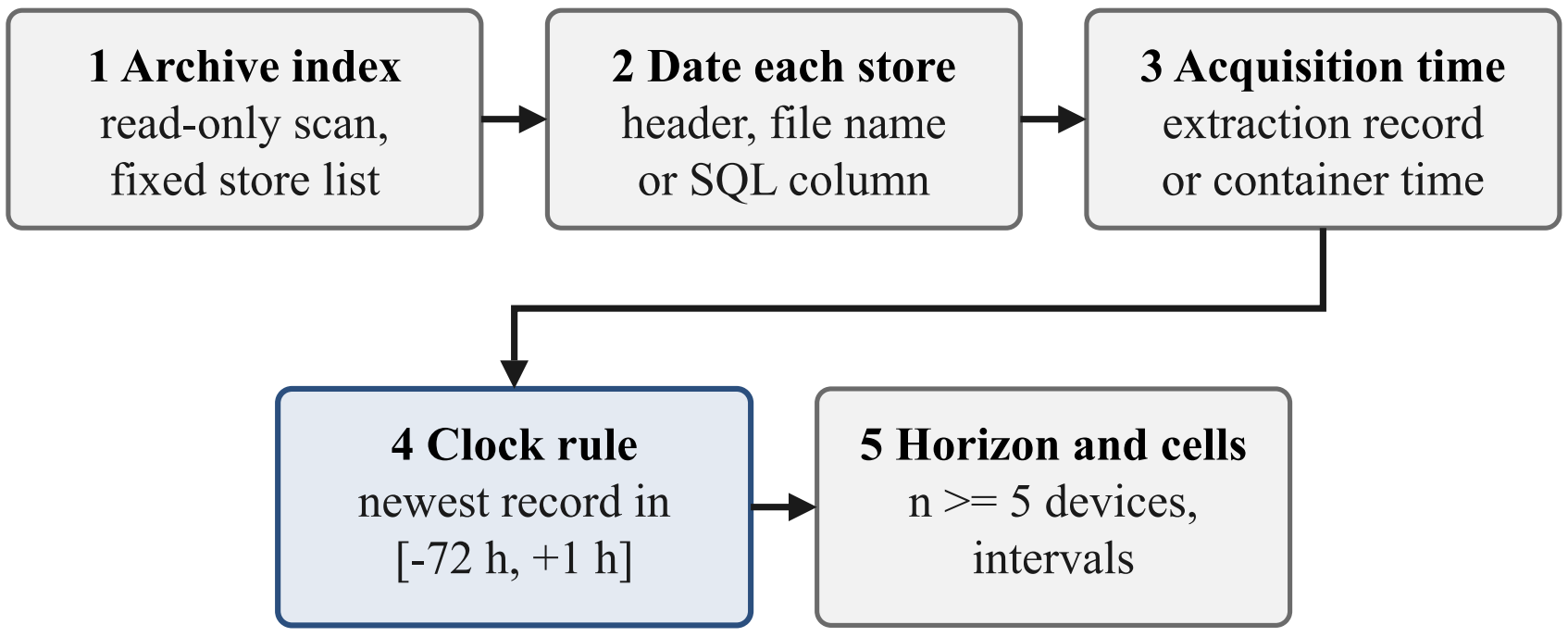

**Fig. 1.** Measurement pipeline. Steps 1 to 3 date each store and the acquisition; step 4 admits a device only if its clock agrees with the acquisition time; step 5 forms horizons and reports cells of at least five devices.

### 3.4 Sampling rule and statistics

Prevalence is reported over every readable extraction. Horizons are reported over devices: one extraction per device (two extractions of one handset made on the same day count once; a comparison of hashed device identifiers found one such pair of iPhone extractions and no other two extractions of one device) and only clock-checked devices. Strata are platform, OS major version where a version has at least five devices (otherwise earlier versions are pooled), and device class where a class has at least five devices. A cell with fewer than five devices is not reported. For each cell we give the median with a distribution-free interval from order statistics (the achieved coverage is at least 95.7% for every interval shown in Tables 4 and 5, and is stated per cell in the runner report), the quartiles, the range, and the share of devices whose horizon reaches 1, 7 and 30 days, with Wilson 95% intervals.

### 3.5 Consistency checks

Three checks ran on every extraction. (1) A name or header time must not be later than the file's modification time plus one hour (plus 14 h for local-time names), because a file is created before it is last written; this tests the reading of names and headers, including the Biome name reading. (2) The oldest SQLite record was compared with and without the write-ahead log. (3) The clock rule and the record filter of Listing 1. As a stratified deep check we looked at the per-stream oldest records of knowledgeC on one iOS 16, one iOS 18 and two iOS 26 devices; what we saw there led to the post hoc analysis of Section 5.7, which we then ran on every device.

## 4 Evidence and Setup

### 4.1 Evidence

The sample frame was every iOS and Android full file system extraction and every Android bug report in the Lab's research corpus that was readable on the day of the study: 34 iOS and 14 Android extractions and seven bug reports (Table 3). An inventory of the collection lists 101 distinct extractions of all kinds; further extractions held elsewhere are not part of this sample. iOS sysdiagnose archives were not analysed, because the readable ones came from three devices, below the cell minimum. The devices were not selected for this study.

The 26 clock-checked Apple devices run iOS 26 (19), iOS 15 to 18 (6) or an undetermined version (1); 18 are iPhones and 8 are Apple devices whose class the archive does not state. The 13 Android phones run Android 16 (9) or Android 10 to 14 (4); 11 are Samsung and 2 come from two other vendors. The seven bug reports come from Android 16 (4) and Android 10 and 15 (3); four are Samsung and three come from three other vendor families. Whether a bug report and an extraction come from the same phone was not established, so the two classes are reported separately.

### 4.2 Setup

One runner script performed every measurement in sequence, each extraction in its own process with a time limit (2,400 s for an iOS extraction, 1,200 s for Android, 900 s for a bug report), and wrote a report of aggregates in Markdown and JSON. It uses only the Python 3.14 standard library: a streaming reader of the archive's central directory that keeps only members of the fixed store list, so that archives of up to several hundred gigabytes are indexed without being unpacked; `zlib` to read the first bytes of a member; and `sqlite3` on scratch copies. All archives were opened read-only. All 55 extractions and reports were read without error; the per-extraction step took 558 s in total.

## 5 Results

### 5.1 Prevalence

Of the 34 iOS extractions, 31 held a dated knowledgeC record, 30 interactionC, 29 each the Persist and Special log streams and powerlog archives, 25 Biome App.InFocus segments, 11 the Signpost stream and 2 the HighVolume stream. The extractions without unified logs came from archives that did not contain the diagnostics folder, and one before-first-unlock extraction held none of the stores. Of the 14 Android extractions, all held daily usage statistics, network statistics and process statistics, 13 DropBox entries, 11 battery history files (all Samsung), and 9 notification history files. All seven bug reports held both log sections.

### 5.2 iOS horizons

Horizons on iOS spread over three orders of magnitude (Table 4, Fig. 2). The Persist stream of the unified log reached a median of 2.5 days and passed one week on 6 of 25 devices; the Signpost stream reached a median of 7.2 days, powerlog archives 7.0 days and the Special stream 24.0 days. Biome App.InFocus segments reached back at least 29.3 days on every device. knowledgeC reached a median of 157 days, but that value comes from one stream (Section 5.7). interactionC reached 166 to 170 days on 18 of 25 devices, a regularity that suggests an age limit of about 24 weeks; of the other seven, five lay below this band (four on iOS 26) and two above it, both on releases before iOS 26.

### 5.3 Android horizons and the source limits

Android stores stayed within the limits in the platform source (Table 5). Daily usage statistics never exceeded 10.2 days, and 10 of 13 phones lay between 9 and 10.2 days, at the ten-day prune of the source [9]; the three shorter horizons occurred on phones whose usage files of other intervals were older than 30 days, so a short daily horizon did not indicate a new device. DropBox entries never exceeded 3.1 days (default three days [5]), and notification history never exceeded 1.5 days (one day [7]). Network statistics lay between 75 and 105 days on 10 of 13 phones, consistent with 15-day files deleted after 90 days [6]; two phones held older files, one of them in the legacy directory beside the newer module directory (manual inspection). Process statistics, bounded by a file count rather than an age [8], reached a median of 3.2 days but up to 102 days. Usage files of any interval reached a median of 331 days, which reflects the long-interval aggregate files, not daily detail.

### 5.4 Bug-report logs

The log sections of a bug report reached back hours. The event log never covered more than 0.121 days (2.9 h) on any of the seven reports, and no event log began with stray lines. The system log began with stray lines in four reports (three Samsung, one other vendor): 1, 3, 39 and 287 lines, each group separated by a gap of 7.3 hours to 16.5 days from a dense body of 0.19 to 3.1 hours. Counted from the oldest line, the system log reached at least one day on three of the seven reports (up to 16.6 days, median 0.31 days), all three from Samsung phones; the oldest line was 39 to 247 times older than the start of the body. Counted from the start of the body (Table 5), no system log reached back more than 0.131 days (3.1 h), the median was 0.04 days (1.0 h) and none reached one day. The stray lines are real log lines. On the four reports they were mostly lines of a few rarely writing components, with some crash-handler and runtime-exception entries, which fits a part of the log that is overwritten much more slowly than the busy part; we did not establish which buffer holds them. They show that such an entry can survive for days, but they do not show that the rest of the log did. The start of the body was the same for gap thresholds from 5 to 60 minutes and stray fractions from 0.1% to 2%; at 5% the rule began to cut into the body of one or two reports. The number of system-log lines, from 4,329 to 405,850, did not order the body horizons (Spearman $\rho = 0.54$, $n = 7$; 0.36 from the oldest line). No report showed a backward step of more than one hour in either section.

**Listing 1.** Clock rule and record filter, fixed before the run.

```
D = newest device-clock time, measured stores
T = acquisition time (record, else file time)
clock-checked   if  T - 72 h <= D <= T + 1 h
record dropped  if  t < 2007-01-01 or t > T + 1 h
bug report flagged if a backward step > 1 h
```

## 5.5 Variation within one release

Within iOS 26 and within Android 16, horizons of stores bounded by size or by a file count varied by more than an order of magnitude, while stores bounded by age varied little (Table 6). The Persist stream on iOS 26 ranged from 0.39 to 21.2 days. The size of the Persist folder did not explain this: its median was 414.7 MB (quartiles 188.7 to 465.3 MB, maximum 518.8 MB, n = 25), and size and horizon were nearly uncorrelated (Spearman ρ = 0.09). The implied write rate ranged from 5.4 to 727.8 MB per day (median 129.2), which is consistent with a store kept to a size, whose span depends on how fast the device writes. interactionC on iOS 26 had quartiles of 167.0 and 168.1 days and never exceeded 169.7 days; daily usage statistics on Android 16 had quartiles of 9.2 and 9.9 days.

**Table 3.** Sample, acquisition-time source and clock rule. “Record” is the extraction record's end time; “file” is the archive's modification time.

| Class | Extr. | Devices | Time source (devices) | Clock-checked |
|---|---|---|---|---|
| iOS full file system | 34 | 33 | record 18, file 15 | 26 |
| Android full file system | 14 | 14 | record 11, file 3 | 13 |
| Android bug report | 7 | 7 | report header | 7 |

Cells for earlier releases are small. Pooled iOS 15 to 18 cells of five or six devices gave medians of 1.6 days for Persist, 31.6 days for Special and 168.6 days for interactionC; no single earlier iOS release or Android release other than 16 reached five devices, and no device class other than iPhone and Samsung did.

## 5.6 Which stores cover 30 days

Fig. 3 shows, per store, the share of devices whose horizon reached 30 days. On iOS, interactionC did on all 25 devices and Biome App.InFocus on 20 of 22; the Special log stream did on 11 of 25 and the Persist stream on 1 of 25. On Android, network statistics did on 12 of 13 and Samsung battery history on 5 of 11; daily usage statistics, DropBox entries and notification history did on none. No bug-report log did. At least one store reached 30 days on 25 of 26 Apple devices and 12 of 13 Android phones, but on iOS those stores were interaction and app-focus records, not the system log, and on Android they were network totals and long-interval usage aggregates, not daily detail.

## 5.7 One stream sets the knowledgeC horizon (post hoc)

The stratified deep check showed that the oldest knowledgeC record belonged to the stream `/discoverability/signals` on all four devices examined, with horizons from 113 to 730 days, while the other streams of the same databases ended much earlier. Run on every clock-checked device, the same query found this stream to be the oldest on 22 of 26. Without it and without two bookkeeping streams for knowledge synchronisation, the knowledgeC horizon had a median of 28.6 days (95% CI 28.0 to 36.2; quartiles 28.0 to 37.9) instead of 157.1, and reached 30 days on 11 of 26 devices instead of 25. The median of the per-stream horizons on a device was 28.2 days across devices. The four-week statement of the earlier literature [10], [13] therefore describes most streams of these databases, and the oldest record of the database does not.

### 5.8 Clocks and consistency

Table 7 summarises the checks. Every `tracev3` header time and every Biome segment name with a usable file time lay before that file's modification time, which supports both readings. For Android names the check held on all files except 13 of 245 usage-statistics files, whose names lie more than an hour after their modification time; we did not find the cause. Applying the write-ahead log never changed the oldest knowledgeC record, but on 18 of 25 devices the interactionC main file alone held an older oldest record than the database with its log applied (median difference over the 25 devices 1.2 days, at most 112.0 days): the log already held deletions of old rows that had not been copied into the main file. A reader that ignores the log would report a longer horizon than the device's database holds.

Of the seven devices that failed the clock rule, six had only the archive's file time as the acquisition reference; for these the gap measures the archive's later handling, not the device clock. The seventh, a tablet with an extraction record, had written its newest record 471 days before acquisition, which fits a device left unused as well as a wrong clock. One iOS device had every store under one day, consistent with a device set up shortly before acquisition; it passed the rule and remains in the sample.

## 6 Discussion

### 6.1 Horizons follow the mechanism that bounds a store

The results separate three kinds of store. Stores pruned by age sat at their limit on most devices: Android daily usage statistics at ten days, DropBox entries at up to three days, notification history at about one day, network statistics at 75 to 105 days, and interactionC at about 168 days. For these, a stated limit predicts the horizon of a device that has run long enough, and the Android source was a reliable guide. Stores kept to a size or a file count did not have a typical horizon: the unified log Persist stream and Android process statistics varied by more than an order of magnitude within one release, because their span depends on how much the device writes. For these, a single published figure, such as hours to ten days [4], is a statement about one device. The third kind mixes the two: the Special stream of the unified log ended near 30 days on many devices, but much earlier on others, consistent with entries purged by type [16].

### 6.2 The oldest record is not the usable window

A horizon computed from a store's oldest record can overstate the usable window in four ways. A single long-lived stream can set it, as `/discoverability/signals` did for knowledgeC. A coarse aggregate can set it, as yearly usage files did for Android usage statistics. And a database read without its write-ahead log can show rows the device has already deleted, as interactionC did. And a few stray old lines can sit above the body of a log, as in four of the seven bug-report system logs, where the oldest line was up to 247 times older than the start of the body. The window an examiner can use is the horizon of the stream or table that answers the question, read with its log applied. The newest lag matters as well: the newest Biome segment was started a median of 9.7 days before acquisition and the newest powerlog archive 3.1 days before, because recent data sit in a current file that the archive horizon does not describe.

**Table 4.** iOS retention horizons in days, clock-checked devices. Median with order-statistic 95% interval; reach 7 d and 30 d as devices (Wilson 95% interval for 30 d). HighVolume (2 devices) is not reported.

| Store | n | Median [95% CI] | Q1–Q3 | Min–max | ≥7 d | ≥30 d [CI] |
|---|---|---|---|---|---|---|
| Unified log, Persist | 25 | 2.5 [1.3, 6.4] | 1.2–6.8 | 0.39–48.2 | 6 | 1 [1%, 20%] |
| Unified log, Signpost | 9 | 7.2 [5.0, 17.6] | 5.4–8.5 | 0.54–48.1 | 6 | 1 [2%, 44%] |
| Unified log, Special | 25 | 24.0 [14.8, 30.2] | 14.1–30.2 | 0.11–70.7 | 20 | 11 [27%, 63%] |
| Powerlog archives | 24 | 7.0 [6.2, 10.5] | 6.2–10.8 | 5.9–32.3 | 12 | 2 [2%, 26%] |
| Biome App.InFocus | 22 | 36.8 [31.4, 68.2] | 31.4–64.7 | 29.3–165.2 | 22 | 20 [72%, 98%] |
| knowledgeC, all records | 26 | 157.1 [82.3, 258.8] | 73.5–270.5 | 0.52–730.0 | 25 | 25 [81%, 99%] |
| knowledgeC, other streams[a] | 26 | 28.6 [28.0, 36.2] | 28.0–37.9 | 0.52–419.9 | 25 | 11 [26%, 61%] |
| interactionC | 25 | 168.0 [167.0, 168.1] | 167.0–168.3 | 49.0–557.5 | 25 | 25 [87%, 100%] |

[a] Post hoc: all streams except `/discoverability/signals` and the knowledge-sync bookkeeping streams (Section 5.7).

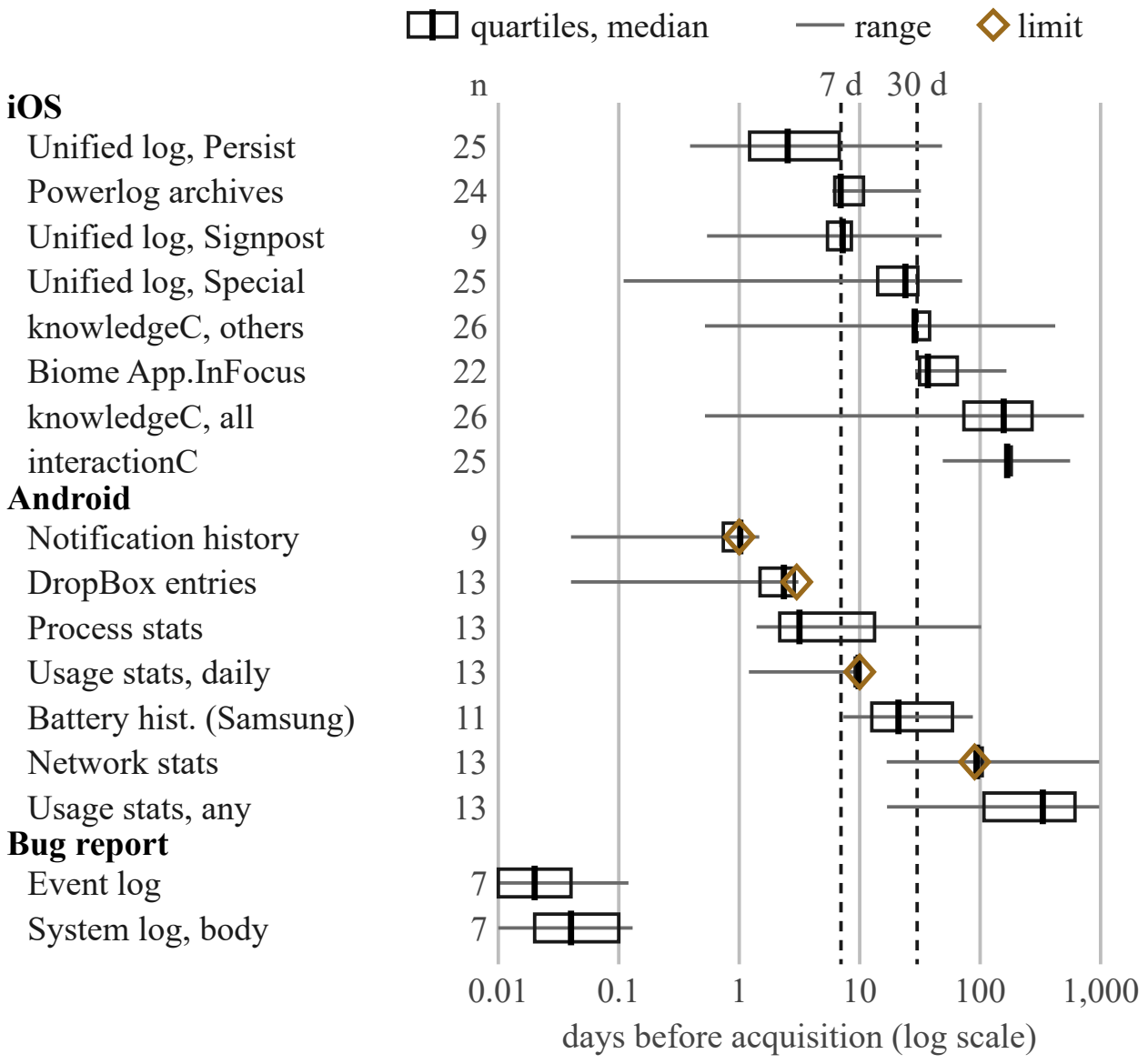

**Fig. 2.** Retention horizon per store on clock-checked devices (log scale; n at left). Boxes show quartiles and the median, lines the range; diamonds mark the limit stated in the Android source. Dashed lines mark 7 and 30 days. The bug-report system log is counted from the start of its body. Source: Tables 4 and 5.

**Table 5.** Android and bug-report horizons in days, clock-checked devices, with the limit stated in the source. "Within" counts devices at or under the stated bound. Bug-report logs are counted from the start of the body of the section (Section 3.1).

| Store | n | Median [95% CI] | Q1–Q3 | Min–max | Source limit | Within |
|---|---|---|---|---|---|---|
| Usage statistics, daily | 13 | 9.8 [2.4, 10.0] | 9.2–9.9 | 1.2–10.1 | 10 d | 13 (≤10.2) |
| Usage statistics, any interval | 13 | 330.8 [102.8, 756.1] | 107.5–614.8 | 16.9–977.0 | yearly files, 2 y | – |
| Notification history | 9 | 1.0 [0.09, 1.1] | 0.74–1.0 | 0.04–1.5 | 1 d | 9 (≤1.5) |
| DropBox entries | 13 | 2.4 [1.2, 3.0] | 1.5–2.9 | 0.04–3.1 | 3 d default | 13 (≤3.1) |
| Network statistics | 13 | 94.2 [79.2, 104.5] | 90.7–104.0 | 16.8–977.0 | 90 d + 15 d file | 11 (≤105) |
| Process statistics | 13 | 3.2 [1.8, 28.5] | 2.2–13.3 | 1.4–102.1 | 8 files | – |
| Battery history (Samsung) | 11 | 21.0 [10.1, 70.3] | 12.6–59.0 | 7.3–87.0 | none found | – |
| Bug report, system log (body) | 7 | 0.04 [<0.01, 0.13] | 0.02–0.10 | <0.01–0.13 | buffer size | – |
| Bug report, event log (body) | 7 | 0.02 [<0.01, 0.12] | 0.01–0.04 | <0.01–0.12 | buffer size | – |

**Table 6.** Horizons within one release, days: median, quartiles and range.

| Store | n | Median | Q1–Q3 | Min–max |
|---|---|---|---|---|
| ***iOS 26*** | | | | |
| Persist | 19 | 2.5 | 1.2–6.6 | 0.39–21.2 |
| Special | 19 | 22.9 | 13.3–30.1 | 0.11–31.9 |
| Powerlog archives | 19 | 7.1 | 6.2–11.0 | 5.9–30.3 |
| Biome App.InFocus | 19 | 35.5 | 31.2–61.2 | 29.3–165.2 |
| knowledgeC, all records | 19 | 156.9 | 65.6–210.8 | 31.9–698.1 |
| interactionC | 19 | 167.9 | 167.0–168.1 | 49.0–169.7 |
| ***Android 16*** | | | | |
| Usage statistics, daily | 9 | 9.8 | 9.2–9.9 | 1.5–10.1 |
| DropBox entries | 9 | 2.7 | 1.5–2.9 | 0.04–3.1 |
| Notification history | 8 | 1.0 | 0.58–1.1 | 0.04–1.5 |
| Network statistics | 9 | 91.9 | 79.2–102.8 | 16.8–104.5 |
| Process statistics | 9 | 3.2 | 2.2–13.3 | 1.7–46.2 |
| Battery history (Samsung) | 8 | 19.5 | 12.4–56.0 | 7.3–70.3 |

## 6.3 What this means for the order of work

On these devices, the body of the system log of a bug report held 3.1 hours at most, and the Persist stream of the unified log days at most, often hours, while app-focus and interaction records held weeks to months. A question about the last day is answerable from many stores; a question about the month before an event depends on a small set of stores and on reading them correctly. Any tool or examiner can report a store's oldest record, its newest lag and whether a write-ahead log was applied, so that the window of each answer is stated with it.

## 7 Threats to Validity

**Internal validity.** Each store was dated by one field. The unified-log header time is the start of a file, so the oldest record may lie slightly after it; Biome segment names were read as creation times, supported by 102 of 102 checks against file times but not by documentation. Powerlog and two Android names carry only a day or a local time; the resulting uncertainty (one day, 14 h) is small against the reported horizons. The cause of 13 usage-statistics name exceptions was not found. The analysis of Section 5.7 is post hoc: it was added after the stratified deep check, although it was then run on every device. The horizon of bug-report logs is counted from the start of the body, not from the oldest line, because a few stray old lines can sit above the body (Section 3.1). This rule is post hoc: the stray lines were found in a Lab study of gaps in logs, the rule was written down before it was applied, and the oldest-line values are given alongside. The ten-minute and 0.5% constants are a choice, tested in Section 5.4 over a range, not derived. The explanation of the stray lines is a reading of their tags and counts, not a test.

**Construct validity.** A horizon measures what a device still held, not what the store can hold. A recently set-up device, a device switched off for a time, or a reset bounds the horizon from below the store's limit; we did not measure device age, except through Android usage files of long intervals. The clock rule bounds the clock error at acquisition to 72 h; it does not detect an earlier clock change that later corrected itself, and it excludes devices that had stopped writing as well as devices with wrong clocks. Nine of the 26 clock-checked Apple devices and two of the 13 Android phones have only an archive file time as acquisition reference; they passed the rule, so the time agreed with their newest record to within the rule's window.

**External validity.** The sample is the Lab's research corpus, not a random sample of phones. Nineteen of 26 Apple devices run iOS 26, 9 of 13 Android phones run Android 16, and 11 of 13 are Samsung; earlier releases and other vendors appear only in pooled cells or not at all. Usage patterns of the users of these devices shape the size-bounded horizons. Results are stated per cell and are not generalised to other populations.

**Conclusion validity.** Cells are small; intervals for the median are wide where n is near five, and a share of 0 of 19 still allows up to 17%. Correlations are reported only as descriptions.

**Limitations.** We did not measure sysdiagnose archives (three devices), Screen Time (covered in LAB-TR-0009), App Ops histories (LAB-TR-0003), notification stores on iOS, the Android intrusion log, or stores of third-party apps. We did not read record content, so we cannot say how complete a store is inside its horizon. The study measured horizons; it does not establish when a given record was deleted.

## 8 Conclusion and Future Work

How far back a phone can answer a question depends on the store, and public statements on this rest on single devices. Horizons ran from hours to years. Stores pruned by age sat at their limits: Android daily usage statistics, DropBox entries and notification history within 10.2, 3.1 and 1.5 days, interactionC near 168 days. Stores kept to a size did not have a typical horizon: the unified log Persist stream ranged from 0.39 to 21.2 days within iOS 26. The oldest record misled when one knowledgeC stream outlived the others, when interactionC was read without its write-ahead log, and when stray old lines sat above the body of a bug-report log, which reached back 3.1 hours at most. On these devices, a horizon belongs to a store, a release and a device; it can be measured from the extraction itself, and any reader can report it with its results.

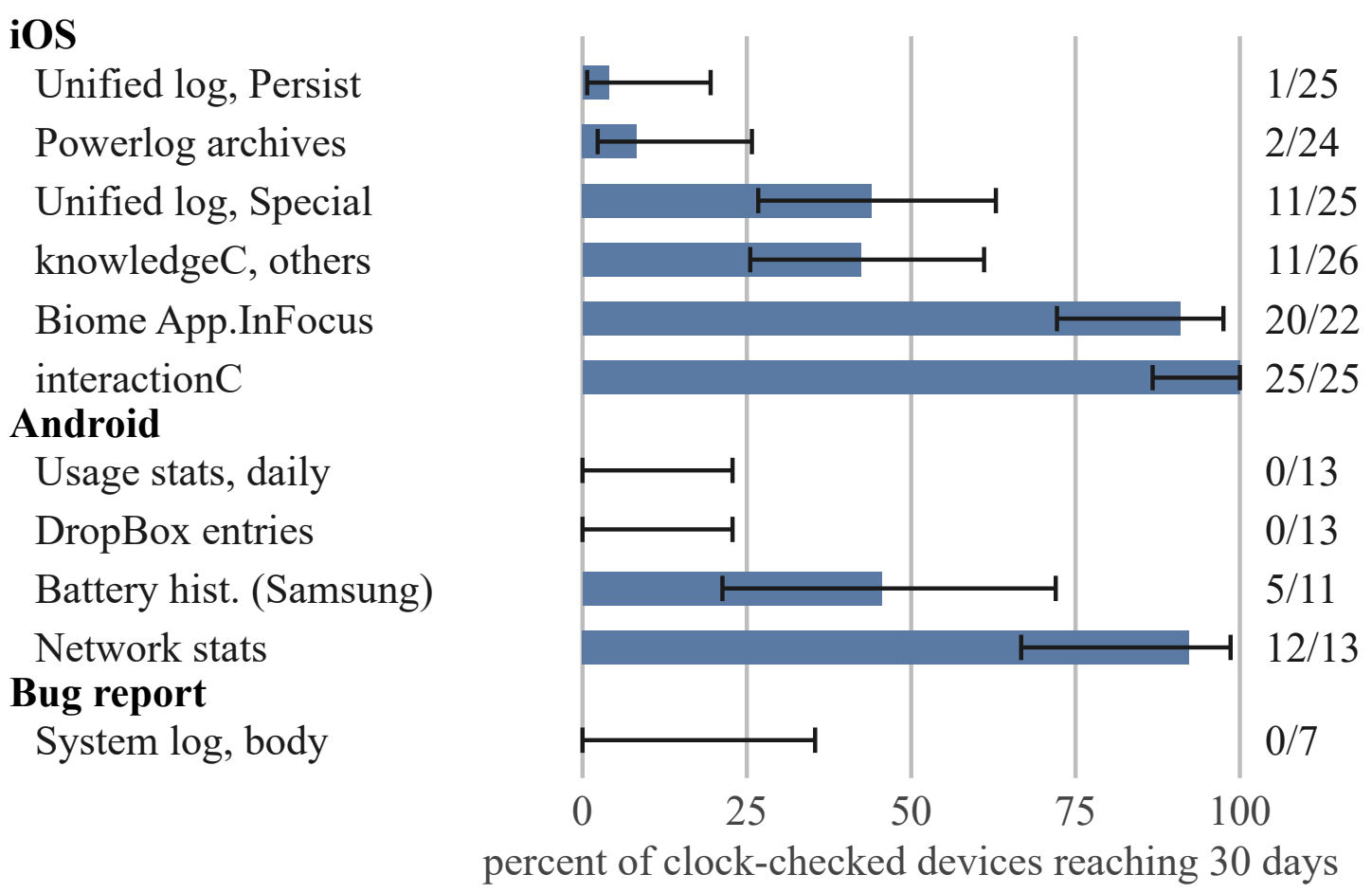

**Fig. 3.** Share of clock-checked devices whose store reached back 30 days, with Wilson 95% intervals; counts at right. The bug-report system log is counted from the start of its body. Source: Tables 4 and 5.

**Table 7.** Consistency and clock checks over all readable extractions (clock rule over devices).

| Check | Result |
|---|---|
| Header time ≤ file time + 1 h (unified log) | Persist 923/923, Special 10,759/10,759, Signpost 635/635, HighVolume 8/8 files |
| Name time ≤ file time + 1 h (Biome) | 102/102 segments with a usable file time |
| Name time ≤ file time (Android; +14 h for local-time names) | DropBox 3,625/3,625; battery 293/293; network 239/239; process 120/120; notification 94/94; usage 232/245 |
| Oldest record changed by the write-ahead log | knowledgeC 0/26; interactionC 18/25, always older without the log |
| Clock rule failed | iOS 6 of 32 devices with an anchor (5 with only an archive file time); Android 1 of 14 (archive file time) |
| Newest record before acquisition, passing devices | iOS median 1.8 h (0.31 to 43.05 h, n = 26); Android median 0.92 h (0.3 to 35.09 h, n = 13) |
| Records excluded by the filter | 7 unified-log headers dated before 2007 (3 extractions); 1,314 rows dated after acquisition (interactionC 1,294 in 4 extractions, knowledgeC 20 in 2) |
| Bug report: backward step > 1 h | 0 of 7 reports in either section |

**Future work**, in order of value: (1) the same protocol on the larger pool of extractions the Lab holds, so that earlier releases and other vendors reach the cell minimum; (2)

controlled runs on the Lab's own test devices with known usage, to separate a store's limit from the device's write rate and age; (3) per-stream horizons for Biome and the unified log, and the sysdiagnose archives once five devices are available.

### Artifact Availability

The readers, the analysis code and a generator of invented archives are released as a package (Python 3.10 or later, standard library only) at https://github.com/danjethh/sectratek-lab-artifacts/tree/main/reports/lab-tr-0021. Code is under the MIT licence, data and text under CC BY 4.0. From the values printed in this report the package rebuilds Table 2, Fig. 1, the intervals of the 30-day column of Table 4, Figs. 2 and 3, and the checks of Listing 1 and of the interval coverage. It also runs the whole pipeline on small invented archives, including a write-ahead log built by hand and bug reports with stray head lines; that run recovers the invented horizons and shows how the code works, but it does not reproduce the findings. Tables 3 to 7 and the numbers of Section 5 were computed on the Lab's research corpus, which cannot be released, so they cannot be reproduced from the package. The package does not contain the Lab's census, its per-extraction process control with time limits, or a reader of the extraction record: the acquisition time is given in a configuration file. The extractions and bug reports cannot be released.

### Data Handling Statement

The evidence for this study is drawn from the Lab's research corpus of mobile device extractions and bug reports and was read only. The report gives aggregate counts and device classes. It contains no identifiers, personal data, message or file content, or case information. Only aggregate statistics, counts, OS versions and device classes are reported; no names, numbers, accounts, identifiers, record content, absolute dates, file paths of the collection or per-device values appear in this paper or its figures.